\documentclass[fleqn,10pt]{wlscirep}
\title{Quantum Rabi Model with Trapped Ions}

\author[1,*]{J. S. Pedernales}
\author[2]{I. Lizuain}
\author[1]{S. Felicetti}
\author[3]{G. Romero}
\author[1]{L. Lamata}
\author[1,4]{E. Solano}
\affil[1]{Department of Physical Chemistry, University of the Basque Country UPV/EHU, Apartado 644, 48080 Bilbao, Spain}
\affil[2]{Department of Applied Mathematics, University of the Basque Country UPV/EHU, Plaza Europa 1, 20018 Donostia-San Sebastian, Spain}
\affil[3]{Departamento de F\'isica, Universidad de Santiago de Chile (USACH), Avenida Ecuador 3493, 917-0124, Santiago, Chile}
\affil[4]{IKERBASQUE, Basque Foundation for Science, Maria Diaz de Haro 3, 48013 Bilbao, Spain}

\affil[*]{julensimon@gmail.com}

\begin{abstract}
We propose the quantum simulation of the quantum Rabi model in all parameter regimes by means of detuned bichromatic sideband excitations of a single trapped ion. We show that current setups can reproduce, in particular, the ultrastrong and deep strong coupling regimes of such a paradigmatic light-matter interaction. Furthermore, associated with these extreme dipolar regimes, we study the controlled generation and detection of their entangled ground states by means of adiabatic methods. Ion traps have arguably performed the first quantum simulation of the Jaynes-Cummings model, a restricted regime of the quantum Rabi model where the rotating-wave approximation holds. We show that one can go beyond and experimentally investigate  the quantum simulation of coupling regimes of the quantum Rabi model that are difficult to achieve with natural dipolar interactions.  
\end{abstract}
\begin{document}

\flushbottom
\maketitle
% * <john.hammersley@gmail.com> 2015-02-09T12:07:31.197Z:
%
%  Click the title above to edit the author information and abstract
%
\thispagestyle{empty}

%\noindent Please note: Abbreviations should be introduced at the first mention in the main text ? no abbreviations lists. Suggested structure of main text (not enforced) is provided below.

\section*{Introduction}

The quantum Rabi model (QRM) describes the most fundamental light-matter interaction involving quantized light and quantized matter.  It is different from the Rabi model~\cite{Rabi36}, where light is treated classically. In general, the QRM is used to describe the dipolar coupling between a two-level system and a bosonic field mode. Although it plays a central role in the dynamics of a collection of quantum optics and condensed matter systems~\cite{FocusCcQEDNJPhys15}, such as cavity quantum electrodynamics (CQED), quantum dots, trapped ions, or circuit QED (cQED), an analytical solution of the QRM in all coupling regimes has only recently been proposed~\cite{Braak11,SolanoPhysics11}. In any case, standard experiments naturally happen in the realm of the Jaynes-Cummings (JC) model~\cite{Jaynes63}, a solvable system where the rotating-wave approximation (RWA) is applied to the QRM. Typically, the RWA is valid when the ratio between the coupling strength and the mode frequency is small. In this sense, the JC model is able to correctly describe most observed effects where an effective two-level system couples to a bosonic mode, be it in more natural systems as CQED~\cite{OpticalCQEDreview05,GarchingCQEDreview06,ParisCQEDbook13}, or in simulated versions as trapped ions~\cite{NISTreview03,InnsbruckReview08} and cQED~\cite{Wallraff04,cQEDreview13}. However, when the interaction grows in strength until the ultrastrong coupling (USC)~\cite{Ciuti05,Niemczyk10,FornDiaz10} and deep strong coupling (DSC)~\cite{Casanova10,DeLiberato14} regimes, the RWA is no longer valid. While the USC regime happens when the coupling strength is some tenths of the mode frequency, the DSC regime requires this ratio to be larger than unity. In such extreme cases, the intriguing predictions of the full-fledged QRM emerge with less intuitive results.

Recently, several systems have been able to experimentally reach the USC regime of the QRM, although always closer to conditions where perturbative methods can be applied or dissipation has to be added. Accordingly, we can mention the case of circuit QED~\cite{Niemczyk10, FornDiaz10}, semiconductor systems coupled to metallic microcavities~\cite{Todorov09, Askenazi14, Kena-Cohen13}, split-ring resonators connected to cyclotron transitions~\cite{Scalari12}, or magnetoplasmons coupled to photons in coplanar waveguides~\cite{Muravev11}. The advent of these impressive experimental results contrasts with the difficulty to reproduce the nonperturbative USC regime, or even approach the DSC regime with its radically different physical predictions~\cite{Werlang08,DeLiberato09,Casanova10,Stassi13,Wolf13,Felicetti14,DeLiberato14}. Nevertheless, these first achievements, together with recent theoretical advances, have put the QRM back in the scientific spotlight. At the same time, while we struggle to reproduce the subtle aspects of the USC and DSC regimes, quantum simulators~\cite{Feynman82} may become a useful tool for the exploration of the QRM and related models~\cite{Ballester12, Mezzacapo14}.

Trapped ions are considered as one of the prominent platforms for building quantum simulators~\cite{InnsbruckReviewQSim12}. In fact, the realization and thorough study of the JC model in ion traps, a model originally associated with CQED, is considered a cornerstone in physics~\cite{HarocheNobel13,WinelandNobel13}.  This is done by applying a red-sideband interaction with laser fields to a single ion~\cite{Diedrich89,Blockley92,NISTreview03} and may be arguably presented as the first quantum simulation ever implemented. In this sense, the quantum simulation of all coupling regimes of the QRM in trapped ions would be a historically meaningful step forward in the study of dipolar light-matter interactions. In this article, we propose a method that allows the access to the full-fledged QRM with trapped-ion technologies by means of a suitable interaction picture associated with inhomogeneously detuned red and blue sideband excitations. Note that, in the last years, bichromatic laser fields have been successfully used for different purposes~\cite{SorensenMolmer99,Solano99,Solano01,Haljan05}. In addition, we propose an adiabatic protocol to generate the highly-correlated ground states of the USC and DSC regimes, paving the way for a full quantum simulation of the experimentally elusive QRM.

\section*{Results}

Single atomic ions can be confined using radio-frequency Paul traps and their motional quantum state cooled down to its ground state by means of sideband cooling techniques~\cite{InnsbruckReview08}. In this respect, two internal metastable electronic levels of the ion can play the role of a quantum bit (qubit). Driving a monochromatic laser field in the resolved-sideband limit allows for the coupling of the internal qubit and the motional mode, whose associated Hamiltonian reads ($\hbar=1$)
 \begin{equation}
 \nonumber
 H  =  \frac{\omega_0}{2} \sigma_z + \nu a^\dag a + \Omega(\sigma^+ +  \sigma^-) \Big({\rm exp}\{i[\eta(a + a^\dag) - \omega_lt + \phi_l]\} +  {\rm exp}\{-i[\eta(a + a^\dag) -\omega_l t + \phi_l]\} \Big) .
\end{equation}
Here, $a^\dag$ and $a$ are the creation and annihilation operators of the motional mode, $\sigma^+$ and $\sigma^-$ are the raising and lowering Pauli operators, $\nu$ is the trap frequency, $\omega_0$ is the qubit transition frequency, $\Omega$ is the Rabi coupling strength, and $\eta=k\sqrt{\frac{\hbar}{2m\nu}}$ is the Lamb-Dicke parameter, where $k$ is the component of the wave vector of the laser on the direction of the ions motion and $m$ the mass of the ion~\cite{NISTreview03}; $\omega_l$ and $\phi_l$ are the corresponding frequency and phase of the laser field. For the case of a bichromatic laser driving, changing to an interaction picture with respect to the uncoupled Hamiltonian, ${H_0 = \frac{\omega_0}{2}\sigma_z} + \nu a^\dag a $ and applying an optical RWA, the dynamics of a single ion reads~\cite{NISTreview03}
\begin{eqnarray}
H^{\rm I} & = & \!\! \sum_{n=r,b}\frac{\Omega_n}{2}\left[e^{i\eta[a(t)+a^\dag(t)]}e^{i(\omega_0-\omega_n)t}\sigma^++\textrm{H.c.}\right] ,
\label{trapped_ion_hamil}
\end{eqnarray}
with  $a(t)=ae^{-i\nu t}$ and $a^\dag (t)=a^\dag e^{i\nu t}$. 
We will consider the case where both fields are off-resonant, first red-sideband (r) and first blue-sideband (b) excitations, with detunings $\delta_r$ and $\delta_b$, respectively,
\begin{eqnarray}
\omega_r = \omega_0 - \nu + \delta_r , \,\,\, \omega_b = \omega_0 + \nu + \delta_b . \nonumber
\end{eqnarray}
In such a scenario, one may neglect fast oscillating terms in Eq.~(\ref{trapped_ion_hamil}) with two different vibrational RWAs. We will restrict ourselves to the Lamb-Dicke regime, that is, we require that $\eta\sqrt{\langle a^\dag a\rangle}\ll1$. This allows us to select terms that oscillate with minimum frequency, assuming that weak drivings do not excite higher-order sidebands, $\delta_n,\Omega_n\ll\nu$ for $n = r , b$. These approximations lead to the simplified time-dependent Hamiltonian
\begin{equation}
\label{RabiInteraction}
\bar{H}^{\rm I} = \frac{i\eta\Omega}{2} \sigma^+ \left( a e^{-i\delta_rt} + a^\dag e^{-i\delta_bt} \right) + \textrm{H.c.},
\end{equation}
where we consider equal coupling strengths for both sidebands, $\Omega = \Omega_r = \Omega_b$.
Equation~\ref{RabiInteraction} corresponds to the interaction picture Hamiltonian of the Rabi Hamiltonian with respect to the uncoupled Hamiltonian $H_0 = \frac{1}{4}(\delta_b+\delta_r)\sigma_z + \frac{1}{2}(\delta_b-\delta_r) a^\dag a $, 
\begin{equation}
H_{\rm QRM} = \frac{\omega_0^R}{2}\sigma_z + \omega^R a^\dag a + ig (\sigma^+-\sigma^-)(a+a^\dag) ,
\label{effective_hamiltonian}
\end{equation} 
with the effective model parameters given by
\begin{eqnarray}
\omega_0^R=-\frac{1}{2}(\delta_r+\delta_b), \,\, \omega^R = \frac{1}{2}(\delta_r-\delta_b), \,\, g=\frac{\eta\Omega}{2} ,
\end{eqnarray}
where the qubit and mode frequencies are represented by the sum and difference of both detunings, respectively. The tunability of these parameters permits the study of all coupling regimes of the QRM via the suitable choice of the ratio $ g / \omega^R$. It is important to note that all realized interaction-picture transformations, so far, are of the form $\alpha a^\dag a +\beta\sigma_z$. This expression commutes with the observables of interest, $\{ \sigma_z, | n \rangle \langle n |, a^{\dagger} a \}$, warranting that their experimental measurement will not be affected by the transformations.

\subsection*{Accessible regimes} The quantum Rabi model in Eq.~(\ref{effective_hamiltonian}) will show distinct dynamics for different regimes, which are defined by the relation among the three Hamiltonian parameters: the mode frequency $\omega^R$, the qubit frequency $\omega_0^R$, and the coupling strength $g$. 

\begin{figure}[ht]
\centering
\includegraphics[width=\linewidth]{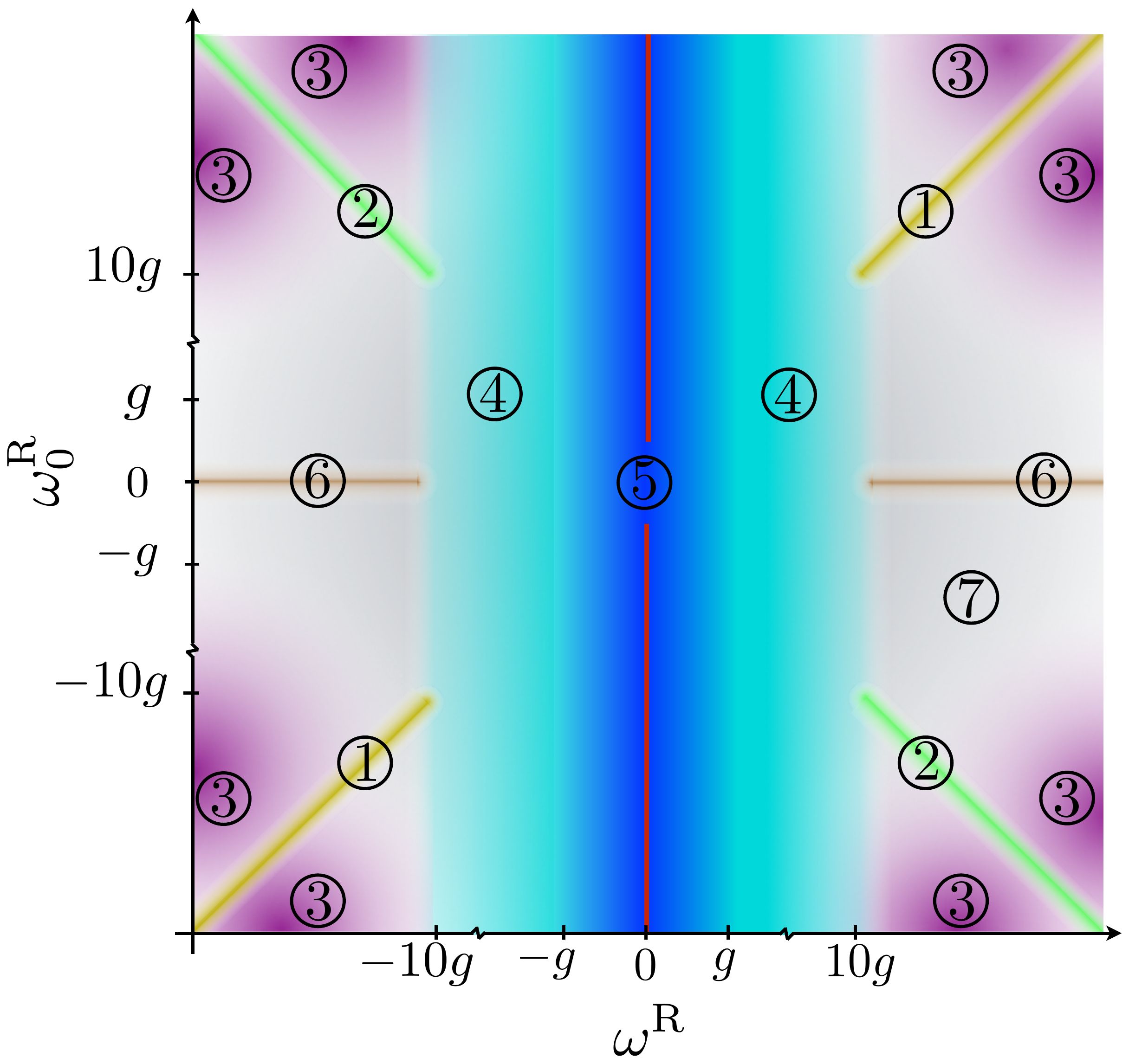}
\caption[]{Configuration space of the QRM. 
(1) JC regime: $ g \ll \{|\omega^R|,|\omega_0^R| \}$ and $| \omega^R - \omega_0^R | \ll |\omega^R + \omega_0^R | $.
(2) AJC regime: $ g \ll \{|\omega^R |,|\omega_0^R| \}$ and $|\omega^R - \omega_0^R| \gg | \omega^R + \omega_0^R | $.
(3) Two-fold dispersive regime: $ g < \{ | \omega^R | , | \omega_0^R | , | \omega^R - \omega_0^R |, |\omega^R + \omega_0^R | \}$.
(4) USC regime: $| \omega^R | < 10 g $, 
(5) DSC regime: $ | \omega^R | < g $, 
(6) Decoupling regime: $ | \omega_0^R | \ll g \ll |\omega^R| $.
(7) This intermediate regime ($ |\omega_0^R| \sim g \ll |\omega^R|$) is still open to study.
The (red) central vertical line corresponds to the Dirac equation regime. Colours delimit the different regimes of the QRM, colour degradation indicates transition zones between different regions. All the areas with the same colour correspond to the same region.}
\label{rabiregions}
\end{figure}

We first explore the regimes that arise when the coupling strength is much weaker than the mode frequency $g \ll |\omega^R| $. Under such a condition, if the qubit is close to resonance, $|\omega^R| \sim |\omega_0^R|$, and {$|\omega^R + \omega_0^R| \gg | \omega^R - \omega_0^R |$} holds, the RWA can be applied. This implies neglecting terms that in the interaction picture rotate at frequency $ \omega^R + \omega_0^R $, leading to the JC model. This is represented in Fig.~\ref{rabiregions} by the region 1 in the diagonal. Notice that these conditions are only possible if both the qubit and the mode frequency have the same sign. However, in a quantum simulation one can go beyond conventional regimes and even reach unphysical situations, as when the qubit and the mode have frequencies of opposite sign. In this case, $| \omega^R - \omega_0^R| \gg | \omega^R + \omega_0^R | $ holds, see region 2, and we will be allowed to neglect terms that rotate at frequencies $ | \omega^R - \omega_0^R |$. This possibility will give rise to the anti-Jaynes Cummings (AJC) Hamiltonian, $H^{AJC} = \frac{\omega_0^R}{2} \sigma_z + \omega^R a^\dag a + ig (\sigma^+a^\dag - \sigma^-a)$. It is noteworthy to mention that, although both JC and AJC dynamics can be directly simulated with a single tuned red or blue sideband interaction, respectively, the approach taken here is fundamentally different. Indeed, we are simulating the QRM in a regime that corresponds to such dynamics, instead of directly implementing the effective model, namely the JC or AJC model.

If we depart from the resonance condition and have all terms rotating at high frequencies $ \{ |\omega^R|, |\omega^R_0|, |\omega^R + \omega_0^R |, | \omega^R - \omega_0^R | \} \gg g$, see region 3, for any combination of frequency signs, the system experiences dispersive interactions governed by a second-order effective Hamiltonian. In the interaction picture, this Hamiltonian reads
\begin{equation}
H_{\textrm{eff}}  =  g^2\left[\frac{|e\rangle\langle e|}{\omega^R - \omega_0^R}- \frac{|g\rangle\langle g| }{\omega^R + \omega_0^R}\right. + \left.\frac{2 \omega^R }{(\omega_0^R + \omega^R)(\omega^R - \omega_0^R)}a^\dag a\sigma_z\right],
\end{equation}
inducing AC-Stark shifts of the qubit energy levels conditioned to the number of excitations in the bosonic mode.

The USC regime is defined as $0.1 \lesssim g/\omega^R \lesssim 1$, with perturbative and nonperturbative intervals, and is represented in Fig.~\ref{rabiregions} by region 4. In this regime, the RWA does not hold any more, even if the qubit is in resonance with the mode. In this case, the description of the dynamics has to be given in terms of the full quantum Rabi Hamiltonian. For $g/\omega^R \gtrsim 1$, we enter into the DSC regime, see region 5 in Fig.~\ref{rabiregions}, where the dynamics can be explained in terms of phonon number wave packets that propagate back and forth along well defined parity chains~\cite{Casanova10}.

In the limit where $\omega^R=0$, represented by a vertical centered red line in Fig.~\ref{rabiregions}, the quantum dynamics is given by the relativistic Dirac Hamiltonian in 1+1 dimensions, 
\begin{equation}
H_D=mc^2\sigma_z+cp\sigma_x,
\end{equation}
which has been successfully implemented in trapped ions~\cite{Gerritsma10, Lamata07}, as well as in other platforms~\cite{Salger11, Dreisow10}.

Moreover, an interesting regime appears when the qubit is completely out of resonance and the coupling strength is small when compared to the mode frequency, $ \omega^R_0 \sim 0 $ and $ g \ll |\omega^R| $. In this case, the system undergoes a particular dispersive dynamics, where the effective Hamiltonian becomes a constant. Consequently,  the system does not evolve in this region that we name as decoupling regime, see region 6 in Fig.~\ref{rabiregions}. The remaining regimes correspond to region 7 in Fig.~\ref{rabiregions}, associated with the parameter condition $ |\omega_0^R| \sim g \ll |\omega^R|$.

The access to different regimes is limited by the maximal detunings allowed for the driving fields, which are given by the condition $\delta_{r,b} \ll \nu$, ensuring that higher-order sidebands are not excited. The simulations of the JC and AJC regimes, which demand detunings $|\delta_{r,b}| \le | \omega^R | + |\omega_0^R| $, are the ones that may threaten such a condition. We have numerically simulated the full Hamiltonian in Eq.~(\ref{trapped_ion_hamil}) with typical ion-trap parameters: $\nu=2\pi \times 3 {\rm MHz}$, $\Omega=2\pi \times 68 {\rm kHz}$ and $\eta=0.06$~\cite{Gerritsma10}, while the laser detunings were $\delta_b=- 2\pi \times 102 {\rm kHz}$ and $\delta_r=0$, corresponding to a simulation of the JC regime with $g/\omega^{\rm R}=0.01$. The numerical simulations show that second-order sideband transitions are not excited and that the state evolution follows the analytical JC solution with a fidelity larger than $99\%$ for several Rabi oscillations. This confirms that the quantum simulation of these regimes is also accessible in the lab. We should also pay attention to the Lamb-Dicke condition $\eta \sqrt{\langle a^\dag a \rangle} \ll 1$, as evolutions with an increasing number of phonons may jeopardize it. However, typical values like $\eta=0.06$ may admit up to some tens of phonons, allowing for an accurate simulation of the QRM in all considered regimes.

Regarding coherence times, the characteristic timescale of the simulation will be given by $t_{\rm char}= \frac{2\pi}{g}$. In our simulator, $g=\frac{\eta \Omega}{2}$, such that $t_{\rm char}=\frac{4 \pi}{\eta \Omega}$. For typical values of $\eta=0.06-0.25$ and of $\Omega/2\pi=0-500$~kHz, the dynamical timescale of the system is of milliseconds, well below coherence times of qubits and motional degrees of freedom in trapped-ion setups~\cite{InnsbruckReview08}. 

\subsection*{State preparation} 
\begin{figure}[ht]
\centering
\includegraphics[width=\linewidth]{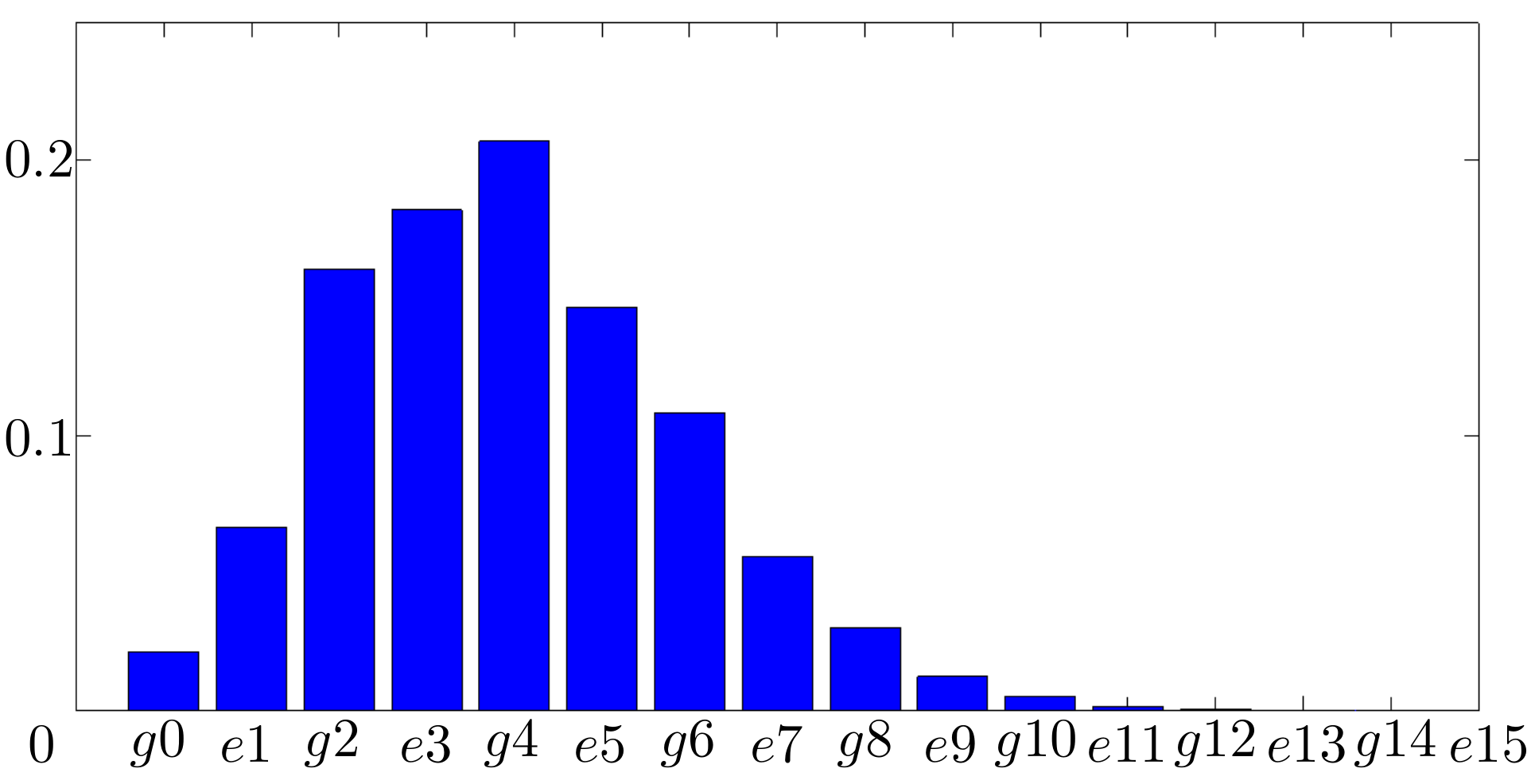}
\caption[]{State population of the QRM ground state. We plot the case of $g/\omega^R=2$, parity $p=+1$, and corresponding parity basis $\lbrace |g,0\rangle,|e,1\rangle,|g,2\rangle,|e,3\rangle,\hdots\rbrace$. Here, $p$ is the expectation value of the parity operator ${P=\sigma_z e^{-i\pi a^\dag a}}$~\cite{Casanova10}, and only states with even number of excitations are populated. We consider a resonant red-sideband excitation ($\delta_r=0$), 
a dispersive blue-sideband excitation ($\delta_b/2\pi=-11.31$kHz), and $g=-\delta_b$, leading to the values $\omega^R=\omega_0^R=-\delta_b/2$ and $g/\omega^R=2$.}
\label{Gstate_barplot_fig}
\end{figure}

The ground state $|G\rangle$  of the QRM in the JC regime ($g\ll\omega^R$) is given by the state $|g,0\rangle$, that is, the qubit ground state, $| g \rangle$, and the vacuum of the bosonic mode, $| 0 \rangle$. It is known that $|g,0\rangle$ will not be the ground state of the interacting system for larger coupling regimes, where the contribution of the counter-rotating terms becomes important~\cite{hwang2010}. As seen in Fig.~\ref{Gstate_barplot_fig}, the ground state of the USC/DSC Hamiltonian is far from trivial~\cite{Braak11}, essentially because it contains qubit and mode excitations, $\langle G|a^\dag a|G\rangle >0$.

\begin{figure}[ht]
\centering
\includegraphics[width=\linewidth]{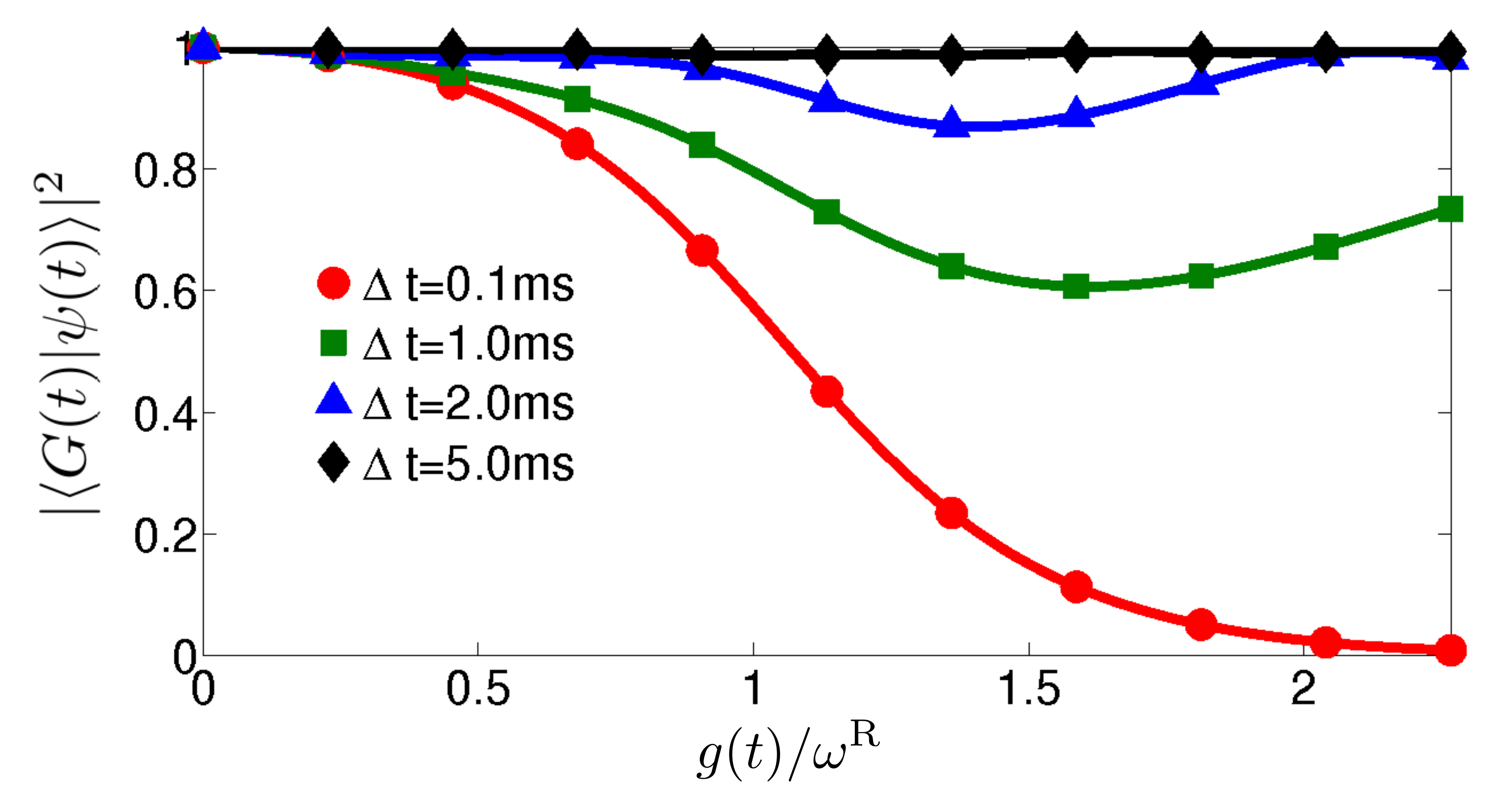}
\caption[]{Fidelity of the adiabatic evolution. Let us assume that the system is initially prepared in the JC ground state $|g,0\rangle$, that is, when $g\ll\omega^R$.  Then, the coupling is linearly chirped during an interval $\Delta t$ up to a final value $g_f$, i. e., $g(t)=g_f t/\Delta t$. For slow changes of the laser intensity, the ground state is adiabatically followed, whereas for non-adiabatic processes, the ground state is abandoned. The instantaneous ground state $|G(t)\rangle$ is computed by diagonalizing the full Hamiltonian at each time step, while the real state of the system $|\psi(t)\rangle$ is calculated by numerically integrating the time dependent Schr\"odinger equation for a time-varying coupling strength $g(t)$.
For the simulation, a $^{40}$Ca$^+$ ion has been considered with parameters: 
$\nu=2\pi\times3$MHz, $\delta_r=0$, $\delta_b=-6\times10^{-4}\nu$, $\eta=0.06$ and $\Omega_f=2\pi\times68$kHz~\cite{Gerritsma10}.}
\label{JC_groundstate_adiabatic_fig}
\end{figure}

Hence, preparing the qubit-mode system in its actual ground state is a rather difficult state-engineering task in most parameter regimes, except for the JC limit. The non analytically computable ground state of the QRM has never been observed in a physical system, and its generation would be of theoretical and experimental interest. We propose here to generate the ground state of the USC/DSC regimes of the QRM via adiabatic evolution. Figure \ref{JC_groundstate_adiabatic_fig} shows the fidelity of the state prepared following a linear law of variation for the coupling strength at different evolution rates. When our system is initialized in the JC region, achieved with detunings $\delta_r=0$ and $|\delta_b |\gg g$, it is described by a JC Hamiltonian with the ground state given by $|G\rangle=|g,0\rangle$. Notice that the $g/\omega^{\rm R}$ ratio can be slowly turned up, taking the system adiabatically through a straight line in the configuration space to regions 4-5~\cite{Kyaw2014}. This can be done either increasing the value of $g$ by raising the intensity of the driving, or decreasing the value of $\omega^{\rm R}$ by reducing the detuning $|\delta_b|$. The adiabatic theorem~\cite{messiah} ensures that if this process is slow enough, transitions to excited states will not occur and the system will remain in its ground state.  As expected, lower rates ensure a better fidelity. 

Once the GS of the QRM is generated, one can extract the populations of the different states of the characteristic parity basis shown in Fig.~(\ref{Gstate_barplot_fig}). To extract the population of a specific Fock state, one would first generate a phonon-number dependent ac-Stark shift~\cite{Solano05}. A simultaneous transition to another electronic state will now have a frequency depending on the motional quantum number. By matching the frequency associated with Fock state $n$, we will flip the qubit with a probability proportional to the population of that specific Fock state. This will allow us to estimate such a population without the necessity of reconstructing the whole wave function.

\section*{Discussion}

We have proposed a method for the quantum simulation of the QRM in ion traps. Its main advantage consists in the accessibility to the USC/DSC regimes and the convenient switchability to realize full tomography, outperforming other systems where the QRM should appear more naturally, such as cQED~\cite{Peropadre10,Felicetti14arXiv}. In addition, we have shown how to prepare the qubit-mode system in its entangled ground state through adiabatic evolution from the known JC limit into the USC/DSC regimes. This would allow for the complete reconstruction of the QRM ground state, never realized before, in a highly controllable quantum platform as trapped ions. The present ideas are straightforwardly generalizable to many ions, opening the possibility of going from the more natural Tavis-Cummings model to the Dicke model. In our opinion, the experimental study of the QRM in trapped ions will represent a significant advance in the long history of dipolar light-matter interactions.

%\section*{Methods}

%Topical subheadings are allowed. Authors must ensure that their Methods section includes adequate experimental and characterization data necessary for others in the field to reproduce their work.

\section*{Acknowledgements}

The authors acknowledge funding from Spanish MINECO FIS2012-36673-C03-01 and FIS2012-36673-C03-02; Ram\'on y Cajal Grant RYC-2012-11391; UPV/EHU Project No. EHUA14/04; UPV/EHU UFI 11/55; UPV/EHU PhD grant; Basque Government IT472-10; Fondecyt 1150653; PROMISCE and SCALEQIT EU projects.

\section*{Author contributions statement}

J.S.P., I.L. and S.F. performed the main calculations and numerical simulations. J.S.P., I.L., S.F., G.R., L.L. and E.S. contributed to the generation and development of the ideas and to the writing of the paper.  

\section*{Additional information}

\textbf{Competing financial interests} The authors declare no competing financial interests. 

\end{document}